# Selective Newsvendor Problem with Dependent Leadtime and Joint Marketing Decisions

Jianing Zhi

School of Business

Jiaxing University

Jiaxing, Zhejiang Province, 314001, China

zhijianing@zjxu.edu

Guanqiu Qi,

Computer Information Systems Department

State University of New York at Buffalo State

Buffalo, NY 14222, USA

qig@buffalostate.edu

Xinghua Li

College of Control Science and Engineering

Zhejiang University

Hangzhou, Zhejiang Province, 310027, China

lixinghua0620@zju.edu.cn

June 23, 2025

# Abstract

In this paper, we investigate a joint decision-making pattern for a two-stage supply chain network, including a supplier, a company, and its customers. We investigate two types of demand patterns, associated with dependent lead time and service level considerations. We define two novel models, including all-or-nothing selective newsvendor problem (AON-SNP) and selective news Vendor Problem with Dependent Lead Time and Price Related Demands (DLSNP). The proposed models are applicable to numerous areas such as the fashion, furniture, and electronic industries. We develop an efficient solution algorithm, referred to as R-search, to identify an optimal solution for the DL-selectivity problem. We examine various responses of the system through parameter sensitivity analysis. Our model proves that if the total demand is lower than the upper limit of the order quantity, the best strategy for $Q$ is to match that demand. If the market increases, more demand comes in, leading to a shortage, and forcing the company to find other local suppliers to fill the additional demand. The results demonstrate that our model well explains various behavior for all involved parties in the network, and provide guidance on intelligent decision making for the company.

# 1 Introduction

Competition in global markets, the introduction of products with shorter life cycles, and heightened customer expectation have driven companies to pay more attention to supply chain operations. Enterprises combine and coordinate different departments such as operations, marketing, and sales to be more responsive and cost-effective. The development of coordination between operations and marketing has been witnessed by both researchers and practitioners since the early 2000s. Explanatory, empirical, and analytical studies between the two business functions are expected to provide the foundation for new frameworks and theories.

In this paper, we investigate a joint management effort between operations and marketing. We consider a newsvendor-like context including a three-level network with two suppliers, a company (buyer) and geographically distributed retailers/customers. We assume that the distant supplier (the primary supplier) has unlimited capacity and that there are no defects or damages during shipping, meaning that the buyer can receive 100% of its orders from the supplier. The second supplier, known as the emergency local supplier, can also provide the same product but with a higher production cost. The company has a certain number of sales agents in the marketing department. The abilities of agents to promote products are different. In addition, each sales agent can handle only a certain number of customers due to limited work hours and energy. We assume that customer demands follow a Poisson distribution. The actual demands are unknown until the sales season begins. After placing an order, each customer has an expected delivery time frame, known as the waiting time; in other words, the waiting time is the number of days a customer is willing to wait. Based on the different orders, the waiting times of customers can vary even for the same product. In the selective distribution network, it often happens that the manufacturer can choose the most appropriate or best-performing outlets and focus effort on them. In an exclusive distribution system, companies usually launch products, such as electronic devices and clothing lines, via exclusive sale or limited edition at selected retailers as test markets. For example, Apple first launched the iPhone 6s in the United States, UK, France, Germany, Canada, Australia, Japan, Hong Kong, and Singapore on September 9, 2015. The second launch wave started on October 9, and the device was available in more than 40 additional countries.

At the beginning of a selling season, the company needs to make two decisions: (i) which customers should be selected to serve and how to assign them to sales agents; (ii) how many products need to order from the supplier. Traditionally, a company makes two decisions separately for marketing and operations. However, conflicts between the two departments will arise when the "supply" of operations does not meet the "demand" of marketing. Specifically, conventional theories of aggregate planning assume that the demand forecast is a mission of marketing and this information flows to operations department to make production and inventory decisions. On the other hand, sales agents in marketing will push for as many orders as they can, without worrying about the capacity, inventory level, and other inside information from the operations. As a consequence, if demands are higher than the capacity of inventory and production, the company may lose profit and customer satisfaction may become worse. In this paper, we propose to integrate decisions of the two functional departments and study how this joint decision-making pattern benefits a company.

Another factor we take into account is lead time. Instead of using a fixed lead time, we consider a lead time that is linearly dependent on the order quantity: the more the company orders, the longer it has to wait. This is a reasonable constraint due to a capacitated supplier. Meanwhile, with a larger order size, the company can serve more customers, but the lead time increases accordingly, which causes the company to lose some customers. In other words, some customers are willing to wait longer for the product delivery, whereas others would take their business elsewhere if the lead time exceeds their waiting times. As such, the company needs to make a trade-off to balance customer satisfaction and profit.

We summarize the contributions of this paper as follows:

- We define two novel models, including all-or-nothing selective newsvendor problem (AON-SNP) and selective newsvendor problem with dependent lead time and price related demands (DL-SNP). We investigate two types of demand patterns, associated with dependent lead time and service level considerations. The proposed models are applicable to numerous areas such as the fashion, furniture, and electronic industries.

- We develop an efficient solution algorithm, referred to as R-Search, to identify an optimal solution for the DL-SNP model. The algorithm employs optimization strategies that are mathematically proven for performance improvement.

- We conduct a series of experiments to validate the DL-SNP model and the R-search algorithm. The results demonstrate that our model well explains various behavior for all involved parties in the network, and provide guidance on intelligent decision making for a company.

The rest of this paper is structured as follows. We deliver a literature review in Section 2, and provide model formulations and related solution approaches in Section 3. In Section 4, we present the results of a series of computational experiments and the marginal insights. We conclude the paper with some key findings and future directions in Section 5.

# 2 Literature Review

In this section, we mainly review prior studies in three areas, including selective newsvendor models, joint decision models of marketing and inventory, and the role of service level constraint in newsvendor problems.

## 2.1 Selective Newsvendor Models

It is common to assume that excess demand is backordered in classic inventory models. However, Gruen et al. (2002) show that only 15% of the customers still decide to place an order for a product that is known to be temporarily out of stock. Thus, most of the excess demand become lost sales rather than backordered in a retail environment. Many prior studies have contributed to the inventory theory regarding lost sales, and we recommend Bijvank and Vis (2011) as a review. Besides, Tsay et al. (1999) provide a review of supply chain contracts and competitive inventory management in a single-period newsvendor setting. In most of the inventory theory papers, the lead time is assumed to be a fixed or a random integral multiple of the review period. Unfortunately, these assumptions are not satisfied in many practical settings. Bijvank and Johansen (2012) consider a periodic review lost sales inventory problem with Poisson demand and constant lead times of any length. Taaffe et al. (2008) consider a newsvendor problem with $n$ potential markets. They develop a model that maximizes the profit, and provide an analysis of the influence of marketing efforts (promotions and advertisements) on the mean and variance of demand. Their model assumes given lead times and normally distributed demand, and also defines a binary variable to capture the market selection decision. Based on the prior model, Chahar and Taaffe (2009) solve a demand selection problem with all-or-nothing orders. They use a conditional value- at-risk approach that allows a decision maker to control the number of demands considered in the overall procurement policy. Also, they formulate a model with the flexibility provided to a firm that wants to balance the risk tolerance versus achievable profits. Khanra et al. (2014) analyze the newsvendor model's robustness, finding conditions linking cost deviation symmetry/skewness to the demand density function. A cost deviation lower bound is set for symmetric unimodal distributions. The newsvendor model proves more sensitive to sub-optimal ordering than the economic order quantity model. Uppari and Hasija (2019) initially develop several newsvendor models capable of theoretically capturing individual-level heterogeneity in order quantities. Subsequently, they employ a comprehensive evaluation approach, integrating theoretical criteria, goodness of fit, and empirical validation, to assess these models and identify the most suitable one. Kou et al. (2025) propose a data-driven stochastic optimization approach to analyze optimal ordering decisions under fluctuating demand and supply uncertainties in the new supply chain paradigm, addressing analytical challenges posed by unknown random variable distributions and balancing shortage and overstocking risks.

In our problem, we use a profit maximization approach based on a newsvendor-type model, in which customers are selected and assigned to sale agents. We consider two demand patterns combined with service level constraints. Besides, we also examine lead time as a linear function of order quantity, which is more reasonable for real world cases. The cost of lost sales (or cost of second local supplier) is also discussed in the model formulation.

## 2.2 Marketing and Inventory Joint Decision Models

Most marketing and inventory integrated models consider price as the primary factor that affects demand quantity and therefore influences order/production quantity or purchasing policies. Researchers have been studying the issue of joint production and pricing decision since the late 1950s. In the literature, most joint production and pricing models focus on the mathematical analysis to determine the optimal pricing and production(or replenishment) planning decisions for a single product over a single period. We recommend (Tang, 2010) as a review of marketing-operations interface models. Note that all previous models assume that the pricing decision can be made after demand (or supply) is realized. Recently, researchers consider other factors that will influence demand such as booking discounts and trade credits. Ouyang et al. (2009) propose an integrated supplier-buyer inventory model with the assumption that market demand is sensitive to the retail price and the payment delay period is dependent on the order quantity. Through an analysis of the total channel profit function, they develop a solution algorithm to determine the optimal order quantity, retail price and the number of shipments from the supplier to the customer. In this study, the supply chain includes a single supplier and a single customer, while our problem considers a single supplier and multiple customers to serve. Carr and Lovejoy (2000) present an inverse newsvendor problem to determine the optimal market selection that maximizes the expected profit, where the order quantity is normally distributed with a known mean and standard deviation. In our paper, both customer selection and order quantity are decision variables. Gao and Hu (2008) develop a joint decision model to concurrently optimize product variety (PV) and inventory control (IC) for multiple product brands in a centralized two-echelon distribution system (dealer and retailer). The model should integrate demand forecasting, analyze the impact of different statistical forecasting methods on decisions and system performance, and be formulated as a nonlinear integer programming model. Its goal is to maximize multi-period profits, considering ordering, shipping, purchasing, inventory holding, and lost sales costs, while adhering to stocking space and service level constraints.

The sales force possess marketing knowledge that is critical for a broad range of decisions. However, researchers have not paid enough attention in the field of marketing-operation joint decisions until recently. Chen (2005) focuses on how incentives affect the sales force and production/inventory decisions under the setting of a single supplier and a single agent. The objective of their model is to maximize a newsvendor profit by determining the level of incentives and order quantity. In their model, customer demands are influenced by three factors: the agent's selling effort, the marketing condition, and random noise. A key assumption is that the firm must make its production decision (order quantity) before observing the total sales. In our problem, we consider the customer selection and assignment based on different sales capabilities of agents, and the two decisions, customer selection and order quantity, are determined at the same time. Also, since the lead time of ordering is linearly dependent on the total quantity, the more the company orders from its supplier, the longer customers have to wait. Customers will leave if the lead time is longer than their waiting times. This characteristic makes our newsvendor model more difficult, which calls for a new solution approach.

## 2.3 Service Level Constraint

In practical situations, it is difficult to measure the shortage cost caused by the loss of potential future customers or orders. Instead of explicitly expressing the shortage cost in the objective function, some studies choose to introduce a service level constraint to reduce the chance of shortage. Aardal et al. (1989) introduce two metrics of service level: the fraction of demand covered from stock and the average number of shortage occurrences per year. They study the relation between shortage cost and service level based on a continuous review inventory system. Chen and Krass (2001) discuss a minimal service level constraint which requires the company to meet a certain service level in each period. Chen and Chuang (2000) propose an extended newsvendor problem by considering an average shortage-level constraint. Their results show that the greater the demand, the earlier an order should be placed. In our model, we utilize the fraction of selected customers and the capacity of each agent as factors to define the service level. Our choice is reasonable since it occurs in practice that only a fraction of customers can be served. Thus, the company can make strategic analyses regarding its profit, reputation, and customer relationship. Jammernegg and Kischka (2013) analyze the integration of financial and non-financial indicators (e.g., customer service, quality, flexibility). Using the newsvendor model, they examine how these potentially conflicting performance metrics influence operational and marketing decisions such as order quantities and pricing, while maintaining the objective of maximizing expected profit. They introduces a service constraint (lower bound for product availability) and a loss constraint (upper bound for loss probability), and provides conditions for the existence of solutions in both models. Abdel-Aal et al. (2017) examine scenarios where selling prices, market entry costs, and product demand patterns vary systematically across markets. Building on this framework, they analyze three distinct formulations of the Multi-Product Selective Newsvendor Problem: flexible market entry, full market entry, and partial market entry. These configurations generate mathematically complex binary nonlinear optimization models.

# 3 Problem Description and Model Formulation

In this section, we formulate two selective newsvendor models based on two demand patterns with the following notations.

*Notation*

**Sets and Indices:**

$\mathcal{I}$ : set of sales agents, $\mathcal{I} = \{i : 1, 2, ...I\}$

$\mathcal{J}$ : set of customers, $\mathcal{J} = \{j : 1, 2, ...J\}$

**Properties of Customer:**

$\mu_j$ : expected demand quantity of customer $j$, $j \in \mathcal{J}$

$w_j$ : waiting time of customer $j$, $j \in \mathcal{J}$

**Properties of Sales Agent:**

$p_{ij}$ : effort of agent $i$ successfully getting an order from customer $j$, $i \in \mathcal{I}$, $j \in \mathcal{J}$

$g_i$ : maximum number of customers which agent $i$, $i \in \mathcal{I}$ can handle

**Time and Costs:**

$a$ : unit production time of supplier

$b$ : fixed shipping time from the supplier to the company

$c$ : production cost

$e$ : salvage price, assume $e < c$

$s$ : shortage cost/second supplier production cost, assume $s > c$

$\lambda$ : sensitivity of customer demand to price reduction

$r$ : base price (suggested price) $\alpha$ :

minimum service level

**Decision Variables:**

$Q$ : order quantity

$R$ : final selling price

$Y_j$ : number of products sold to customer $j$, $j \in \mathcal{J}$

$X_{ij} = \begin{cases} 1, & \text{if customer } j \text{ is assigned to salesman } i \\ 0, & \text{otherwise.} \end{cases}$  $i \in \mathcal{I}, j \in \mathcal{J}$

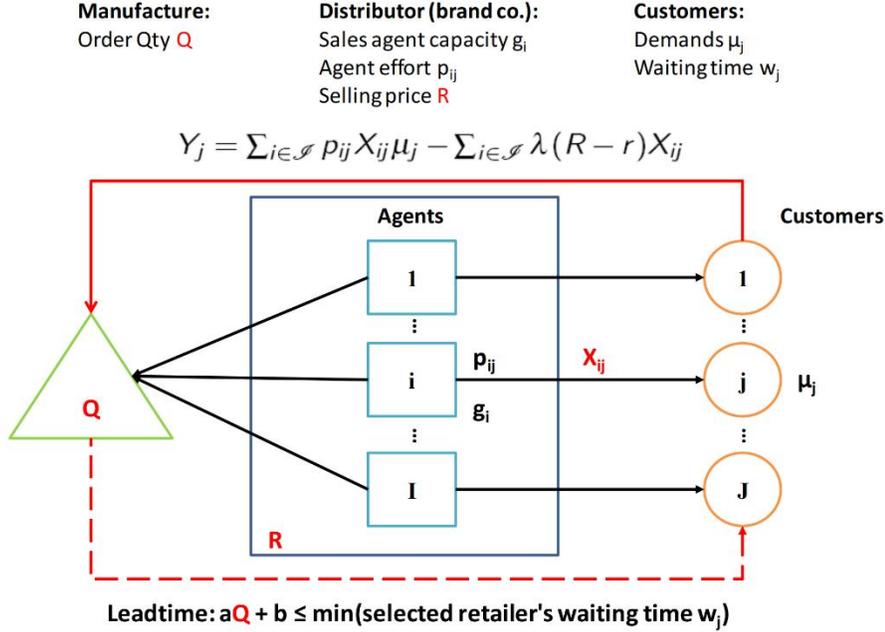

Figure 1: Big Picture of the Selective Newsvendor Network

## 3.1 All-or-Nothing Selective Newsvendor Problem (AON-SNP)

We consider a joint decision model for a two-stage supply chain network, including a supplier, a company, and its customers. The buyer company has a marketing department with a group of sales agents that have different abilities and skills of product promotion. Whether the company can secure customer orders depends on the marketing abilities of agents, which are measured by the success rate to obtain orders. Besides, each agent can only handle a certain number of customers due to limited working hours and personal experience. We assume that customer demands follow a Poisson distribution, and are uncertain until a selling season starts.

Sometimes customers are willing to accept partial orders. However, there are many situations in real life that customers do not accept partial orders, and they simply take the business away if their demands cannot be fully satisfied. This phenomenon usually happens in the retail industry, in particular for the highly customized products, like large conference/wedding purchasing and school uniforms. We refer to this type of customer demand as all-or-nothing (AON) demand. In other words, an order needs to be satisfied at the expected level or not at all.

At the beginning of a selling season, the company desires to maximize the expected profit by making a series of decisions including the strategy of customer selection, a proper selling price, and order quantity from the oversea supplier. A customer has two attributes: the expected demand $\mu_j$, i.e., mean of the Poisson distribution and waiting time $w_j$. Each agent $i$ in the company has influence on order quantities of customers, denoted by $(p_{ij}) \in [0.8, 1.2]$. Also, every agent also has a limited number of customers that he/she can handle, denoted by $g_i$. For the AON-SNP model, the company has two decisions to make: (i) the assignment $X_{ij}$ of customers to each agent, which equals 1 if customer $j$ is assigned to agent $i$ and 0 otherwise; (ii) the order quantity $Q$. The question we attempt to answer is how does this demand pattern influence decision making and the expected profit. As such, we formulate the AON-SNP model as follows.

**Objective (AON-SNP):**

$$\max \quad f(X_{ij}, Q) = r \sum_{j \in \mathcal{J}} \sum_{i \in \mathcal{I}} p_{ij} X_{ij} \mu_j - cQ + e(Q - \sum_{j \in \mathcal{J}} \sum_{i \in \mathcal{I}} p_{ij} X_{ij} \mu_j)$$

**Subject to**:

| | | | |
|---|---|---|---|
| **Agent Capacity** | $\sum_{j \in \mathcal{J}} X_{ij} \leq g_i,$ | $\forall i \in \mathcal{I}.$ | (1) |
| **Customer Assignment** | $\sum_{i \in \mathcal{I}} X_{ij} \leq 1,$ | $\forall j \in \mathcal{J}.$ | (2) |
| **Production sold** | $\sum_{i \in \mathcal{I}} \sum_{j \in \mathcal{J}} p_{ij} X_{ij} \mu_j \leq Q,$ | $\forall i \in \mathcal{I}, \forall j \in \mathcal{J}.$ | (3) |
| **Integrality** | $Q > 0, X_{ij} \in \{0,1\}.$ | | (4) |

Constraint set 1 specifies the capacity for each agent. Constraint set 2 ensures that each customer can either be assigned by one agent or not served at all. In particular, with the AON demand, the number of products sold to a selected customer $j$ equals the demand of $j$. If the company does not have enough inventory as required by customer $j$, he/she will not be served. This constraint leads to a problem that the total fulfilled demand will never exceed the order quantity $Q$, as shown in Constraint 3. Therefore, shortage never occurs. The formulated model is a mixed integer linear programming (MILP) problem, which can be solved by existing MILP solvers such as IBM CPLEX.

## 3.2 Selective Newsvendor Problem with Dependent Lead Time and Price Related Demands(DL-SNP)

The AON-SNP model considers one kind of demand pattern in the real world. However, sometimes customers are flexible to accept partial orders, or companies can use a second supplier to fill the shortage with higher costs. Choosing a second or emergency provider is costly but worthwhile if it can help the company keep loyalty customers. As opposed to the AON-SNP model, this section explores a model that allows the main supplier to take partial order quantities of customers. Moreover, we try to capture the influence of selling price on demands. In particular, we formulate a selective newsvendor model with dependent lead time and price-related demand, as shown in Figure 1. Concretely, the product quantity sold to a customer $j$, i.e., $Y_j$, is negatively affected by the selling price $R$, and a price scale $\lambda$ is introduced to measure the sensitivity of customers to price reduction. For instance, $\lambda = 1.0$ means that if the price rises/reduces by \$1, the demand will decrease/increase by 1 unit, accordingly. Each customer $j$ has an expectation of delivery time $w_j$. When the waiting time is greater than $w_j$, the customer will leave with nothing bought. The waiting time is determined by the production time at supplier, $a \times Q$, and the time consumed on transportation $b$. Therefore, the company needs to make three kinds of decisions to achieve a maximal profit, including 1) the selling price $R$, 2) the order quantity $Q$ from its oversea supplier, and 3) the assignment of a selected customer $j$ to an agent $i$, i.e., $X_{ij}$. To formulate the new model, we first define $Y_j$:

$$Y_j = \sum_{i \in \mathcal{I}} p_{ij} X_{ij} \mu_j - \sum_{i \in \mathcal{I}} \lambda (R - r) X_{ij} \qquad \forall j \in \mathcal{J}$$

We then formulate the DL-SNP model:

**Objective (DL-SNP):**

$$\max \quad f(X_{ij}, Q, R) = \sum_{j \in \mathcal{J}} R Y_j - cQ + e(Q - \sum_{j \in \mathcal{J}} Y_j)^+ - s(\sum_{j \in \mathcal{J}} Y_j - Q)^+ \qquad (5)$$

**Subject to**:

| | | | |
|---|---|---|---|
| **Customer Service Level** | $J - \sum_{i \in \mathcal{I}} \sum_{j \in \mathcal{J}} X_{ij} \leq J(1-\alpha),$ | | (6) |
| **Sales Agent Capacity** | $\sum_{j \in \mathcal{J}} X_{ij} \leq g_i,$ | $\forall i \in \mathcal{I}n$ | (7) |
| **Lead Time** | $aQ + b \leq \sum_{i \in \mathcal{I}} w_j X_{ij} + M(1 - \sum_{i \in \mathcal{I}} X_{ij}),$ | $\forall j \in \mathcal{J}$ | (8) |
| **Customer Assignment** | $\sum_{i \in \mathcal{I}} X_{ij} \leq 1,$ | $\forall j \in \mathcal{J}$ | (9) |
| **Production Sold** | $Y_j = \sum_{i \in \mathcal{I}} p_{ij} X_{ij} \mu_j - \sum_{i \in \mathcal{I}} \lambda (R - r) X_{ij}$ | $\forall j \in \mathcal{J}$ | (10) |
| **Integrality** | $X_{ij} \in \{0,1\},$ | | (11) |
| **Non-negativity** | $Q, R, Y_j \geq 0.$ | | (12) |

As opposed to AON-SNP, we add a few more constraints to meet the needs of the new model. First, we introduce Constraint 6 of customer service level to ensure that a certain fraction of customers will be selected. A minimum service level $\alpha$ is given to enforce the constraint. Second, we add a constraint set 8 to ensure that the lead time should not exceed the expected waiting times of all selected customers. We employ the big-M method to properly formulate this constraint: if customer $j$ is selected, then $\sum_{i \in \mathcal{I}} X_{ij} = 1$, and the constraint becomes $aQ + b \leq \sum_{i \in \mathcal{I}} w_j X_{ij}$, which sets an effective bar for the lead time; otherwise $\sum_{i \in \mathcal{I}} X_{ij} = 0$, and the constraint becomes $aQ + b \leq M$, which holds anyway. The product quantity sold to a customer $j$, $Y_j$, is another key difference. Besides the expected demand and the efforts of a sales agent, we integrate the factors of selling price, base price, and customers' sensitivity to price reduction into the formulation of $Y_j$. In particular, the company can sell more products to a selected customer if 1) the customer has a high demand in the past, indicating a high demand mean, or 2) the assigned agent does a good job to convince the customer to buy more. On the contrary, the customer will order less with a higher selling price $R$.

The DL-SNP model is a mixed integer non-linear problem, which can not be solved by existing solvers such as CPLEX. We develop a solution approach named R-search, which can effectively and efficiently identify an optimal solution.

### 3.2.1 Solution Approach: R-search Method

The proposed R-search method is based on a fact that when $R$ is given, the objective function $f(X_{ij},Q,R)$ becomes $f(X_{ij},Q)$, which is a MILP problem solvable by CPLEX. The key is to determine a proper searching strategy to reduce the running time of the algorithm. Clearly, the larger the searching range is, the more MILP problems need to be solved, which hurts the performance.

To facilitate the presentation of the upcoming lemma and its proof, we define a list of quantities denoted by $K^{max} = \{\max\{p_{ij}\mu_j | i \in \mathcal{I}\} | j \in \mathcal{J}\}$. Essentially, $K^{max}$ can be obtained by selecting the largest value of $p_{ij}\mu_j$ for each customer $j$ and add it into the list. Now let $K^{max}_{sorted}$ denote a sorted $K^{max}$ in ascending order, and $K^{max}_{sorted}[i]$ refer to the $i$th element in $K^{max}_{sorted}$.

**Lemma 1.** *The upper bound of $R$ can be determined as follows:*

$$R_{ub} = \frac{1}{\lambda} K^{max}_{sorted}\left[J - \lceil J\alpha \rceil + 1\right] + r$$

*Proof.* If a customer $j$ is not selected, then $\sum_{i \in \mathcal{I}} X_{ij} = 0$, and $Y_j = 0 \, \Sigma_{i \in I}$, meaning that $j$ has no impact on $R$, as no products are sold to this customer. If $j$ is selected, then $\sum_{i \in \mathcal{I}} X_{ij} = 1$, and we can rewrite the Constraint 10 as:

$$R \leq \frac{\sum_{i \in \mathcal{I}} p_{ij} X_{ij} \mu_j}{\lambda \sum_{i \in \mathcal{I}} X_{ij}} + r$$

$$= \frac{1}{\lambda} \sum_{i \in \mathcal{I}} p_{ij} X_{ij} \mu_j + r$$

$$\leq \frac{1}{\lambda} \max\{p_{ij}\mu_j | i \in \mathcal{I}\} + r$$

Note that $\max\{p_{ij}\mu_j | i \in \mathcal{I}\}$ has been pre-stored in $K^{max}_{sorted}$. To satisfy the service level constraint, at least $\lceil J\alpha \rceil$ customers will be selected. To ensure that Constraint 10 holds for each selected $j$, $R$ is required to be less than the smallest $p_{ij}\mu_j$ among all selected customers. Clearly, that would be the $(J-\lceil J\alpha\rceil+1)$ th element in $K^{max}_{sorted}$. Therefore, we have

$$R \leq R_{ub} = \frac{1}{\lambda} K^{max}_{sorted}\left[J-\lceil J\alpha\rceil+1\right]+r \ . \qquad \square$$

Lemma 1 identifies the upper bound of the selling price $R$, which serves as an accurate starting point for the search. The lemma shows that any price beyond that point is a violation of Constraint 10.

**Lemma 2.** *Without any constraints, the problem can achieve the optimal solution if and only if there is neither a salvage cost nor a shortage cost, i.e., $Q = \sum_{j\in\mathcal{J}} Y_j$.*

*Proof.* Assume $R$ is given, let $\sum_{j\in\mathcal{J}} Y_j = D$ to represent the total product quantity sold, then the objective function becomes

$$f(D,Q) = RD - cQ + e(Q-D)^+ - s(D-Q)^+$$

Since $\min(Q,D) = D-(D-Q)^+ = Q-(Q-D)^+$, we can transform $f(D,Q)$ as:

$$\begin{aligned} f(D,Q) &= RD - c\left(\min(Q,D)+(Q-D)^+\right) + e(Q-D)^+ - s(D-Q)^+ \\ &= RD - c\left(D-(D-Q)^+ + (Q-D)^+\right) + e(Q-D)^+ - s(D-Q)^+ \\ &= (R-c)D - G(D,Q) \end{aligned}$$

in which $G(D,Q) = (s-c)(D-Q)^+ + (c-e)(Q-D)^+$. Now we show that $f$ is maximized if and only if $Q = D$; as such, we have $G(D,Q) = 0$, and $f_{max} = (R-c)D$. There are two cases:

**Case I**: $D > Q$. Without loss of generality, let $D = Q+\Delta, \Delta > 0$ then $f_{max} = (R-c)D = (R-c)(Q+\Delta)$, and $f(Q+\Delta,Q) = (R-c)(Q+\Delta)-(s-c)\Delta > 0$. Therefore, $f_{max} - f(Q+\Delta,Q) = (s-c)\Delta$. In the real world, the shortage cost should be greater than the production cost, i.e., $s > c$, because once shortage occurs, the company will find another supplier to fill the additional demand, which increases the cost. Therefore, instead of spending more money working with another supplier, it is a better to satisfy the additional demand $\Delta$ by the original supplier, which then brings up the order quantity $Q$ by $\Delta$ units so that $D = Q$ once again.

**Case II**: $D < Q$. Similarly, let $D = Q - \Delta$, $\Delta > 0$ then $f_{max} = (R-c)D = (R-c)(Q-\Delta)$, and $f(Q+\Delta, Q) = (R-c)(Q-\Delta) - (c-e)\Delta$. Therefore, $f_{max} - f(Q-\Delta, Q) = (c-e)\Delta > 0$, which also makes sense in the real world, as the salvage price is lower than the production cost, i.e., $e < c$. It is clearly better to reduce the order quantity $Q$ by $\Delta$ units if possible.

In conclusion, the optimality condition for the profit function is that $Q = D$, meaning that neither shortage nor salvage will occur. □

Lemma 2 shows the optimality condition of the profit function without any constraint. In other words, $Q$ and $D$ are free to take any positive values. However, for our problem, $Q$ is limited by the lead time, i.e., $aQ + b$ is less than the shortest waiting time among all selected customers. It is thus possible that when $Q$ reaches its limit, $D$ can still be increased to raise the overall profit, leading to a shortage. To achieve that, $R$ needs to be further reduced to increase the demand. In particular, we formulate the upper limit of $Q$:

**Lemma 3.** *The upper limit of $Q$ is:* $Q_{ub} = \dfrac{\min\{w_j | j \in \mathcal{J}\} - b}{a}$.

*Proof.* To ensure that Constraint 8 is always satisfied, $aQ + b$ is supposed to be less than or equal to the shortest waiting time for all customers, i.e., $aQ + b \leq \min\{w_j | j \in \mathcal{J}\}$. We can then obtain $Q$'s upper limit. □

**Lemma 4.** *When $X_{ij}$ is given, and $Q$ reaches its upper limit, the profit function is concave on $R$.*

*Proof.* There are two cases to discuss:

**Case I:** if $Q = D$, neither shortage nor salvage will occur, and the profit function becomes

$$f(R) = R \sum_{j \in \mathcal{J}} Y_j - cQ$$

$$= R \sum_{j \in \mathcal{J}} \left( \sum_{i \in \mathcal{I}} p_{ij} X_{ij} \mu_j - \sum_{i \in \mathcal{I}} \lambda (R-r) X_{ij} \right) - cQ$$

$$= R \sum_{i \in \mathcal{I}} \sum_{j \in \mathcal{J}} p_{ij} X_{ij} \mu_j - R^2 \lambda \sum_{i \in \mathcal{I}} \sum_{j \in \mathcal{J}} X_{ij} + R \lambda r \sum_{i \in \mathcal{I}} \sum_{j \in \mathcal{J}} X_{ij} - cQ$$

**Case II**: if $Q < D$, there will be a shortage. Similarly, we have

$$f(R) = R\sum_{j\in\mathcal{J}} Y_j - cQ - s\sum_{j\in\mathcal{J}} Y_j + sQ$$

$$= R\sum_{i\in\mathcal{I}}\sum_{j\in\mathcal{J}} p_{ij}X_{ij}\mu_j - R^2\lambda\sum_{i\in\mathcal{I}}\sum_{j\in\mathcal{J}} X_{ij} + R(\lambda r + \lambda s)\sum_{i\in\mathcal{I}}\sum_{j\in\mathcal{J}} X_{ij}$$

$$- s\sum_{i\in\mathcal{I}}\sum_{j\in\mathcal{J}} p_{ij}X_{ij}\mu_j - s\lambda r\sum_{i\in\mathcal{I}}\sum_{j\in\mathcal{J}} X_{ij} + (s-c)Q$$

For both case, we have $f''(R) = -2\lambda < 0$, implying a concave $f$ on $R$. In addition, for both cases, $f_{max}$ is achieved if and only if $f'(R^*) = 0$.

For **case I:** $R^* = \dfrac{\sum_{i\in\mathcal{I}}\sum_{j\in\mathcal{J}} p_{ij}X_{ij}\mu_j}{2\lambda\sum_{i\in\mathcal{I}}\sum_{j\in\mathcal{J}} X_{ij}} + \dfrac{r}{2}$

For **case II:** $R^* = \dfrac{\sum_{i\in\mathcal{I}}\sum_{j\in\mathcal{J}} p_{ij}X_{ij}\mu_j}{2\lambda\sum_{i\in\mathcal{I}}\sum_{j\in\mathcal{J}} X_{ij}} + \dfrac{r+s}{2}$ □

Lemma 4 is a key to reduce the number of evaluations of the MILP prolblem. However, to use it in the solution approach, we need additional theory support.

**Lemma 5.** *If (1) $R$ is given, (2) $Q$ reaches its upper limit, and (3) $Q < D$, then maximizing the profit is equivalent to maximizing $\sum_{i\in\mathcal{I}}\sum_{j\in\mathcal{J}} F_{ij}X_{ij}$, in which $F_{ij} = p_{ij}\mu_j - \lambda(R-r)$.*

*Proof.* If both R and Q are given, and there is a shortage. Then our profit function becomes

$$f(X_{ij}) = (R-s)\sum_{j\in\mathcal{J}} Y_j + (s-c)Q$$

$$= (R-s)\sum_{j\in\mathcal{J}}\left(\sum_{i\in\mathcal{I}} p_{ij}X_{ij}\mu_j - \sum_{i\in\mathcal{I}} \lambda(R-r)X_{ij}\right) + (s-c)Q$$

$$= (R-s)\sum_{j\in\mathcal{J}}\left(\sum_{i\in\mathcal{I}}\left(p_{ij}\mu_j - \lambda(R-r)\right)X_{ij}\right) + (s-c)Q$$

$$= (R-s)\sum_{j\in\mathcal{J}}\sum_{i\in\mathcal{I}} F_{ij}X_{ij} + (s-c)Q$$

Since $R > s$, and $Q = Q_{ub}$, maximizing $f$ is equivalent to maximizing $\sum_{j\in\mathcal{J}}\sum_{i\in\mathcal{I}} F_{ij}X_{ij}$. □

Lemma 5 explains the rationality of choosing new customers: If $F_{ij} > 0$, then customer $j$ should be selected, i.e., $\sum_{i\in\mathcal{I}} X_{ij} = 1$, because a positive $\sum_{i\in\mathcal{I}} F_{ij}X_{ij}$ will contribute to the profit. If, however, $F_{ij} \leq 0$, then customer j will not be selected; thus, $\sum_{i \in \mathcal{I}} F_{ij} X_{ij} = 0$, which does not contribute to the profit. Therefore, the optimizer will select every customer $j$ with a positive $F_{ij}$, as long as there exists an agent who can take new customers.

**Lemma 6.** *During the search, $X_{ij}$ remains unchanged when the number of selected customers does not change.*

*Proof.* Based on Lemma 5, the optimizer will select all customers with a positive $F_{ij}$. In other words, if a customer $j$ is selected, then $F_{ij} = p_{ij}\mu_j - \lambda(R-r) > 0$, i.e., $R < p_{ij}\mu_j/\lambda + r$. Meanwhile, an available agent $i$ with the highest effort $p_{ij}$ will be selected to handle the customer. In conclusion, a customer $j$ will be selected and assigned to an agent $i$ if 1) $R < p_{ij}\mu_j/\lambda + r$, 2) agent $i$ is not overloaded, i.e., $\sum_{j \in \mathcal{J}} X_{ij} \leq g_i$, and 3) $F_{ij}$ is the highest among all agents that can take new customers. However, if there is another agent $i'$ who is also available and has the same effort as agent $i$, the optimizer will break the tie by choosing the agent with a lower agent ID. Compared to Lemma 5, we take one step further. Let us keep reducing $R$, before the next new customer can be selected, $X_{ij}$ will not change. First, all selected customers will remain the same, because unselecting a selected customer will decrease the profit, and given the current $R$, no new customers can be selected. Second, for a selected customer $j$, the optimizer will not assign $j$ to a different agent, say $i'$, with a lower effort $p_{i'j} < p_{ij}$, which lowers the overall expected profit. Therefore, $X_{ij}$ will remain unchanged as long as the number of selected customers does not change. □

Lemma 6 gives an effective indicator of the status of $X_{ij}$. During the search, if the number of selected customers remains unchanged, $X_{ij}$ remains the same, which provides a condition of applying Lemma 4.

Based on the above analysis, we describe the R-search algorithm as follows:

**Step 1**: Determine $R_{ub}$ based on Lemma 1, and use it as the starting point of the search: $R \leftarrow R_{ub}$. Also, let $R_{lb} \leftarrow s$. In the real world, the selling price should be higher than the shortage cost $s$.

**Step 2**: Solve the reduced MILP problem via CPLEX; if the solution is optimal, add tuple ($R$, profit, $Q$, $X_{ij}$) to the result set.

**Step 3**: If $Q$ has not reached its upper limit, i.e., $Q < Q_{ub}$, then $R \leftarrow R - step\_sz$. Go to step 2.

**Step 4**: If $Q$ has reached its upper limit, i.e., $Q = Q_{ub}$:

- If no shortage occurs, i.e., $Q = \sum_{j \in \mathcal{J}} Y_j$. We can calculate a $R'$ based on case I of Lemma 4.

- If shortage occurs, we can calculate a $R'$ based on case II of Lemma 4.

- If $R' > R$, meaning that keep searching downwards will not yield a better solution so the search stops. Go to step 5.

- If the $stop$ condition is set, break the search and go to step 5.

- If $R' < R$, then $R \leftarrow R'$. Meanwhile, if all customers are selected, or all agents are working in full capacity, a stop condition will be set; go to step 2.

**Step 5**: Return the result with the highest profit in the result set.

In addition, we provide the pseudo code of R-search in Algorithm 1.

## 4 Evaluation

We conduct all computational experiments on a workstation with a Intel Core(TM) i7-4770 CPU (3.40GHz), a 16GB RAM, and a 64-bit Windows 10 operation system. The employed solver is IBM CPLEX 12.6. We implement the solution algorithm in C++. In addition, we develop a Python program for data generation.

### 4.1 Data Setting

We present the data setting of the experiment in Table 1. Based on the instance size, we generate two groups of data, each of which consists of 24 distinct instances. Besides the data size, the two data sets only differ in unit production time, as the demand for the large data set is significantly larger than the small set, so we apply a shorter unit production time for the large set. In other words, the company improves its production efficiency to handle more demands.

**Algorithm 1** R-search

**Require:** Parameter list in Table 1, $K_{sorted}^{max}$, step–size

1: rst_set $\leftarrow \emptyset$, $R_{lb} \leftarrow s$, $stop \leftarrow$ False
2: $R_{ub} \leftarrow \frac{1}{\lambda} K_{sorted}^{max} \left[ J - \lceil J\alpha \rceil + 1 \right] + r$ (see Lemma 1)
3: $Q_{ub} \leftarrow \frac{\min\{w_j | j \in \mathcal{J}\} - b}{a}$ (see Lemma 3)
4: $R \leftarrow R_{ub}$
5: **while** $R > R_{lb}$ **do**
6:    $rst \leftarrow Cplex.solve$(reduced DL-SNP)
7:    **if** $rst$ is an optimal solution **then**
8:      Extract tuple (profit, $Q$, $X_{ij}$) from $rst$
9:      Add tuple (profit, $R$, $Q$, $X_{ij}$) to rst_set
10:     Compute $\{Y_j | j \in \mathcal{J}\}$
11:     $D \leftarrow \sum_{j \in \mathcal{J}} Y_j$
12:     **if** $Q < Q_{ub}$ **then**
13:       $R \leftarrow R$ - step_size
14:     **else if** $Q = Q_{ub}$ **then**
15:       **if** $Q = D$ **then**
16:         $R' \leftarrow \frac{\sum_{i \in \mathcal{I}} \sum_{j \in \mathcal{J}} p_{ij} X_{ij} \mu_j}{2\lambda \sum_{i \in \mathcal{I}} \sum_{j \in \mathcal{J}} X_{ij}} + \frac{r}{2}$ (see Lemma 4)
17:       **else if** $Q < D$ **then**
18:         $R' \leftarrow \frac{\sum_{i \in \mathcal{I}} \sum_{j \in \mathcal{J}} p_{ij} X_{ij} \mu_j}{2\lambda \sum_{i \in \mathcal{I}} \sum_{j \in \mathcal{J}} X_{ij}} + \frac{r+s}{2}$ (see Lemma 4)
19:       **end if**
20:       **if** $R' > R$ **or** $stop = true$ **then**
21:         break
22:       **else**
23:         $R \leftarrow R'$
24:       **if** All customers are selected, **or** all agents are in full capacity **then**
25:         Set the stop flag $stop \leftarrow true$
26:       **end if**
27:       **end if**
28:     **end if**
29:    **end if**
30: **end while**
31: Output the result with highest profit in rst_set.

Table 1: Data Specification

| Parameter | Notation | Setting |
|---|---|---|
| # Sales Agents | $I$ | small - [4,6,8,10]; large - [12,14,16,18] |
| # Customers | $J$ | small - [50, 100] by 10; large - [200, 300] by 20 |
| Waiting Time | $w_j$ | U[90, 120] |
| Agent Capacity | $g_i$ | U[20, 40] |
| Mean Demand | $\mu_j$ | U[10, 20] |
| Sales Effort | $p_{ij}$ | U[0.8, 1.2] |
| Unit Prod. Time | $a$ | small - 0.1 day; large - 0.02 day |
| Fixed Shipping Time | $b$ | 3 days |
| Salvage Price | $e$ | $50 |
| Shortage Price | $s$ | $90 |
| Price Scale | $\lambda$ | 1.0 |
| Base Price | $r$ | $100 |
| Production Cost | $c$ | $70 |
| Service Level Limit | $\alpha$ | 0.8 |

## 4.2 Results

We display the results for the small data set and the large data set in table 2 and table 3, respectively. For each instance, we report the optimal profit, the optimal decision variables $R^*$ and $Q^*$, the running time in seconds (denoted by $T$ in the table), the pre-calculated $R_{ub}$, the total actual demand $D = \sum_{j \in \mathcal{J}} Y_j$, and the difference between $Q$ and $D$, denoted by $\Delta_{Q^*,D}$. In addition, to evaluate the demand fulfillment rate, we define three metrics:

$$M_1 = \frac{\sum_{j \in \mathcal{J}} \frac{Y_j}{\mu_j}}{\sum_{i \in \mathcal{I}} \sum_{j \in \mathcal{J}} X_{ij}} \times 100\% \qquad M_2 = \frac{Q}{\sum_{j \in \mathcal{J}} \mu_j} \times 100\% \qquad M_3 = \frac{\sum_{i \in \mathcal{I}} \sum_{j \in \mathcal{J}} X_{ij}}{J} \times 100\%$$

in which $M_1$ computes an average ratio of the actual demand and the expected demand for each individual customer, $M_2$ is a ratio of order quantity fulfilled by the company and total expected demand of the market, and $M_3$ is the actual service level the company achieves. $M_1$ and $M_2$ show the capability of the company fulfilling the market needs from two different aspects.

**Solution:** For both data sets, we obtain the optimal solution for each tested instance. Shortage occurs for 28 out of 48 instances, meaning that in most circumstances the company is unable to fulfill all demands from its dedicated supplier, and has to seek help from a local supplier. We will discuss the potential strategies later in the section of sensitivity analysis. In addition, the achieved service level is 100% for every instance. This means that with our data setting, as long as agents are not in full capacity, the more customers are selected, the more profit the company earns.

Table 2: Results for the Small Data Set

| ID | $I$ | $J$ | Profit | $R^*$ | $Q^*$ | T | $R_{ub}$ | D | $\Delta_{Q^*,D}$ | $M_1$ | $M_2$ | $M_3$ |
|----|----|----|--------|-------|-------|-----|----------|--------|-------|------|------|------|
| 1  | 4  | 50 | 24621  | 98.5  | 863.9 | 2.7 | 116 | 863.9  | 0.0    | 125% | 124% | 100% |
| 2  | 4  | 60 | 29380  | 103.5 | 870   | 2.0 | 118 | 887.4  | -17.4  | 95%  | 94%  | 100% |
| 3  | 4  | 70 | 31816  | 106.0 | 870   | 1.7 | 118 | 898.5  | -28.5  | 82%  | 81%  | 100% |
| 4  | 4  | 80 | 32843  | 105.5 | 870   | 1.5 | 116 | 994.2  | -124.2 | 82%  | 73%  | 100% |
| 5  | 4  | 90 | 35257  | 105.7 | 870   | 1.3 | 116 | 1133.9 | -263.9 | 82%  | 64%  | 100% |
| 6  | 4  | 100| 37411  | 105.8 | 870   | 1.2 | 117 | 1265.3 | -395.3 | 82%  | 57%  | 100% |
| 7  | 6  | 50 | 26164  | 100.0 | 870   | 2.2 | 116 | 876.4  | -6.4   | 118% | 117% | 100% |
| 8  | 6  | 60 | 29967  | 104.5 | 868.6 | 2.0 | 118 | 868.6  | 0.0    | 93%  | 94%  | 100% |
| 9  | 6  | 70 | 31333  | 106.0 | 870   | 1.9 | 117 | 870.8  | -0.8   | 83%  | 84%  | 100% |
| 10 | 6  | 80 | 35139  | 106.6 | 870   | 1.8 | 119 | 1065.5 | -195.5 | 83%  | 69%  | 100% |
| 11 | 6  | 90 | 36145  | 106.1 | 870   | 1.6 | 118 | 1161.8 | -291.8 | 84%  | 64%  | 100% |
| 12 | 6  | 100| 37647  | 105.8 | 880   | 1.2 | 116 | 1266.4 | -386.4 | 84%  | 59%  | 100% |
| 13 | 8  | 50 | 27292  | 101.5 | 866.4 | 2.7 | 119 | 866.4  | 0.0    | 110% | 110% | 100% |
| 14 | 8  | 60 | 29417  | 104.0 | 865.2 | 2.3 | 118 | 865.2  | 0.0    | 96%  | 97%  | 100% |
| 15 | 8  | 70 | 31583  | 105.9 | 870   | 1.7 | 116 | 891.2  | -21.2  | 84%  | 84%  | 100% |
| 16 | 8  | 80 | 33119  | 105.7 | 870   | 1.6 | 115 | 1003.0 | -133.0 | 85%  | 75%  | 100% |
| 17 | 8  | 90 | 36033  | 106.1 | 870   | 1.4 | 116 | 1158.3 | -288.3 | 84%  | 64%  | 100% |
| 18 | 8  | 100| 37229  | 105.7 | 870   | 1.4 | 116 | 1259.5 | -389.5 | 85%  | 60%  | 100% |
| 19 | 10 | 50 | 26294  | 100.5 | 862.1 | 2.6 | 116 | 862.1  | 0.0    | 116% | 116% | 100% |
| 20 | 10 | 60 | 29960  | 104.5 | 868.4 | 2.5 | 118 | 868.4  | 0.0    | 94%  | 95%  | 100% |
| 21 | 10 | 70 | 31575  | 105.9 | 870   | 2.0 | 116 | 891.0  | -21.0  | 85%  | 84%  | 100% |
| 22 | 10 | 80 | 34421  | 106.3 | 870   | 2.1 | 118 | 1043.7 | -173.7 | 84%  | 71%  | 100% |
| 23 | 10 | 90 | 36052  | 106.1 | 870   | 2.1 | 118 | 1158.9 | -288.9 | 85%  | 65%  | 100% |
| 24 | 10 | 100| 38246  | 106.1 | 870   | 1.5 | 116 | 1291.4 | -421.4 | 84%  | 58%  | 100% |

Table 3: Results for the Large Date Set

| ID | $I$ | $J$ | Profit | $R^*$ | $Q^*$ | T | $R_{ub}$ | D | $\Delta_{Q^*,D}$ | $M_1$ | $M_2$ | $M_3$ |
|---|---|---|---|---|---|---|---|---|---|---|---|---|
| 1 | 12 | 200 | 109283 | 96 | 4203 | 7.5 | 118 | 4203 | 0 | 141% | 140% | 100% |
| 2 | 12 | 220 | 120086 | 98 | 4289 | 6.6 | 116 | 4289 | 0 | 130% | 130% | 100% |
| 3 | 12 | 240 | 126885 | 99.5 | 4301 | 6.3 | 117 | 4301 | 0 | 122% | 122% | 100% |
| 4 | 12 | 260 | 138413 | 102 | 4325 | 6.6 | 118 | 4325 | 0 | 108% | 109% | 100% |
| 5 | 12 | 280 | 142468 | 103 | 4317 | 5.5 | 116 | 4317 | 0 | 102% | 103% | 100% |
| 6 | 12 | 300 | 148082 | 104 | 4350 | 5.4 | 116 | 4363 | -13 | 97% | 98% | 100% |
| 7 | 14 | 200 | 109330 | 96 | 4205 | 7.2 | 116 | 4205 | 0 | 142% | 141% | 100% |
| 8 | 14 | 220 | 119641 | 97.5 | 4350 | 6.5 | 116 | 4352 | -2 | 134% | 133% | 100% |
| 9 | 14 | 240 | 128517 | 99.5 | 4350 | 6.9 | 116 | 4370 | -20 | 122% | 122% | 100% |
| 10 | 14 | 260 | 137093 | 101.5 | 4350 | 6.8 | 118 | 4356 | -6 | 111% | 111% | 100% |
| 11 | 14 | 280 | 145249 | 103.5 | 4336 | 6.4 | 118 | 4336 | 0 | 100% | 101% | 100% |
| 12 | 14 | 300 | 149143 | 104 | 4350 | 6.1 | 118 | 4439 | -89 | 97% | 96% | 100% |
| 13 | 16 | 200 | 110677 | 96.5 | 4177 | 8.4 | 118 | 4177 | 0 | 139% | 138% | 100% |
| 14 | 16 | 220 | 120576 | 98 | 4306 | 7.8 | 118 | 4306 | 0 | 131% | 131% | 100% |
| 15 | 16 | 240 | 128584 | 99.5 | 4350 | 7.0 | 116 | 4377 | -27 | 123% | 122% | 100% |
| 16 | 16 | 260 | 137155 | 101.5 | 4350 | 7.1 | 118 | 4361 | -11 | 112% | 112% | 100% |
| 17 | 16 | 280 | 146077 | 103.5 | 4350 | 7.1 | 118 | 4376 | -26 | 101% | 101% | 100% |
| 18 | 16 | 300 | 149012 | 104 | 4350 | 6.9 | 118 | 4429 | -79 | 97% | 97% | 100% |
| 19 | 18 | 200 | 112188 | 96.5 | 4234 | 8.5 | 118 | 4234 | 0 | 139% | 138% | 100% |
| 20 | 18 | 220 | 119787 | 98 | 4278 | 8.3 | 118 | 4278 | 0 | 131% | 131% | 100% |
| 21 | 18 | 240 | 130672 | 100 | 4350 | 8.3 | 118 | 4367 | -17 | 120% | 119% | 100% |
| 22 | 18 | 260 | 140199 | 102.5 | 4314 | 7.3 | 118 | 4314 | 0 | 106% | 107% | 100% |
| 23 | 18 | 280 | 144165 | 103 | 4350 | 7.2 | 118 | 4397 | -47 | 103% | 103% | 100% |
| 24 | 18 | 300 | 147949 | 104 | 4350 | 7.1 | 118 | 4354 | -4 | 97% | 98% | 100% |

**Decision Variables:** For both sets, the optimal selling price $R^*$ is in [$90, $110]; depending on particular instances, $R^*$ could be higher or lower than the suggested selling price $r$. Also, $Q^*$ reaches its upper limit for 32 out of 48 instances.

**Performance:** For the small set, the average running time is 1.87 seconds, while it is 7.03 seconds for the large set. The largest instance we test is ($I$=18, $J$=300) which includes 5402 decision variables in total, and it only takes 7.1 seconds to finish the search. This result validates the effectiveness and efficiency of the R-search algorithm. We also compare the R-search method with a sequential search method in the next subsection.

**Market Satisfaction:** $M_1$ and $M_2$ measure the company's ability to satisfy the market. The difference is that $M_1$ takes into account the additional products from a local supplier when shortage occurs, while $M_2$ only considers Q which is entirely produced by the company's dedicated oversea supplier.

## 4.3 Testing the R-search Algorithm

The proposed R-search algorithm is validated through the experiment. From the small data set, we choose a typical instance ($I$=4, $J$=100) that demonstrates how the search works. We compare the R-search algorithm to a sequential search algorithm, which will search the entire range of $[s, R_{ub}]$ with a step size of 0.5. The results displayed in Figure 2 and 3 show the searching performance for the R-search and the search, respectively. The R-search algorithm finds the optimal solution ($R^* = 105.8$, profit=34711) in 1.23 seconds, while the search algorithm takes 13.85 seconds to find a solution ($R^* = 106$, profit=34708). In terms of the number of MILP problems, the search evaluates 54 MILP problems, while the R-search only evaluates 15 MILP problems, showing a 72% improvement.

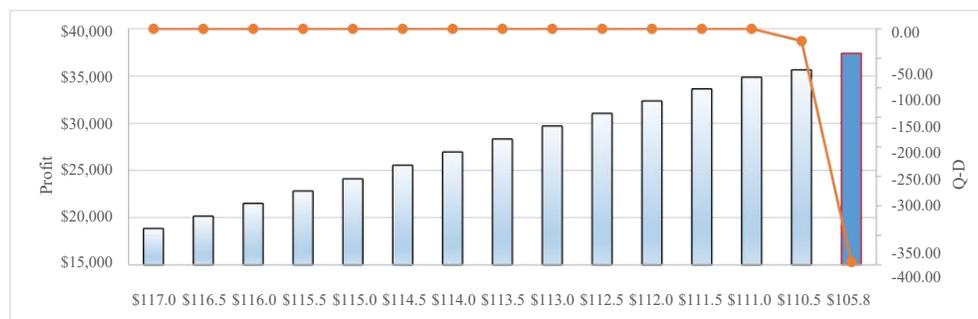

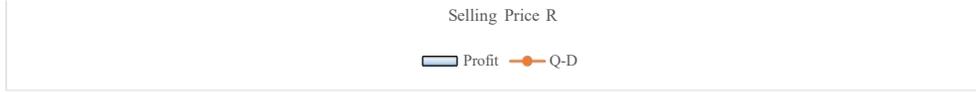

Figure 2: R-search for Instance (*I*=4, *J*=100) with a running time of 1.2 second

The results are also in alignment with our analysis. R-search does a sequential search when there is no shortage, i.e., $Q - D = 0$. Once shortage occurs, and $Q$ has reached its upper limit, R-search will compute a $R^*$ based on lemma 4, which can significantly reduce the number of MILP problems to be evaluated. For this instance, R-search identifies the optimal price $105.8 immediately after the algorithm detects a shortage for the first time ($R = \$110.5$). On the other hand, the sequential search evaluates all MILP problems in the searching range with a fixed step size of 0.5. Figure 3 shows a concave curve for the profit function, although strictly speaking, the profit is only concave on $R$ beyond the point of $R = 110.5$, when shortage occurs and that $X_{ij}$ will not change since all customers are selected.

## 4.4 Parameter Sensitivity Analysis

For the sensitivity analysis on each parameter, we take one instance from the small data set, and only vary the parameter under study, while keeping all other parameters untouched. We list the tested parameters and their values in Table 4.

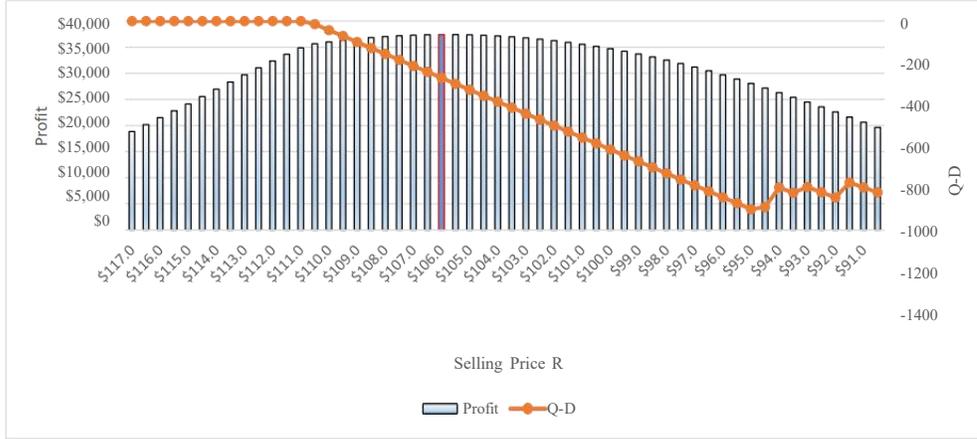

Figure 3: Search for Instance ($I=4$, $J=100$) with a running time of 13.9 seconds

Table 4: Parameter Sensitivity Analysis

| Parameter | Notation | Setting |
|---|---|---|
| Market size | $\mu_j$ | U[8,12], U[10,20], U[15,25], U[30,50] |
| Agent Capacity | $g_i$ | U[5,10], U[10,15], U[15,20], U[20,30] |
| Price Scale | $\lambda$ | [0.1, 1.4] by 0.1 |
| Base Price | $r$ | [$95, $150] by $5 |
| Production Cost | $c$ | [$50, $80] by $5 |
| Unit Production Time | $a$ | [0.1, 0.5] by 0.1 day |
| Service Level Limit | $\alpha$ | [0, 1] by 0.01 |

Table 5: Impact of Market Size on Instance ($I = 4$, $J = 50$)

| Size | $\mu_j$ | Profit | $R^*$ | $Q^*$ | T | $R_{ub}$ | D | $\Delta_{Q^*,D}$ | $M_1$ | $M_2$ | $M_3$ |
|---|---|---|---|---|---|---|---|---|---|---|---|
| XL | U[30, 50] | 62982.8 | 123.8 | 870 | 2.4 | 148 | 1350 | -480 | 66% | 44% | 100% |
| L | U[15, 25] | 32576.3 | 107.5 | 869 | 2.5 | 124 | 869 | 0 | 86% | 87% | 100% |
| M | U[10, 20] | 25682.1 | 99.5 | 870 | 2.4 | 116 | 872 | -2 | 119% | 119% | 100% |
| S | U[8, 12] | 19307.3 | 92.5 | 858 | 2.8 | 112 | 858 | 0 | 176% | 174% | 100% |

### 4.4.1 Market Size

We evaluate four markets, referred to as extra large (XL), large (L), medium (M), and small (S), in terms of the market size. For our problem, the size of a market is determined by the expected demand $\mu_j$. Since the four instances only differ in the market size, the upper limit of $Q$ will be the same, which is an indicator of the production capacity of the company's oversea supplier. The results shown in Table 5 demonstrate that there is a large shortage for the XL market, while for the rest three market no shortage occurs. This is rational because a larger market will force the company to work with the local supplier to fulfill the shortage without extending the deadline of order delivery. However, we notice that even with shortage, the company in the XL market decides to only fulfill two thirds of expected demands, i.e., $M_1 = 66.33\%$, because to match the amount of overall expected demands, the company has to lower the selling price to attract more orders, which might hurt the profit. Our model demonstrates the ability to decide the optimal selling price, striving to fulfill as much demand as possible without losing profit.

### 4.4.2 Agent Capacity

Agent capacity is another factor that determines a company's ability to fulfill demands of a market. To assess its impact, we specify four different levels of agent capacity, including U[5,10], U[10,15], U[15,20], and U[20, 30] and test the setting on instance ($I$=4, $J$=100). In addition, we eliminate the service level constraint by setting $\alpha = 0$, as agent capacity is a hard constraint which might conflict with a pre-set service level, leading to an infeasible solution. We present the results in Table 6. Also, the randomly generated agent capacities for the four instances are [6, 5, 8, 6], [15, 14, 10, 13], [15, 16, 15, 16], and [22, 28, 25, 22]. From $M_3$, we find that for each tested instance, all agents are working in full capacity, which can be validated through $\sum_{i \in \mathcal{I}} g_i = M_3 J$. Therefore, a low $M_3$ due to agent capacity implies that the marketing department is understaffed.

Table 6: Impact of Agent Capacity on Instance ($I$ = 4, $J$ = 100)

| ID | $g_i$ | Profit | $R^*$ | $Q^*$ | T | $R_{ub}$ | D | $\Delta_{Q^*,D}$ | $M_1$ | $M_2$ | $M_3$ |
|---|---|---|---|---|---|---|---|---|---|---|---|
| 1 | U[5,10] | 16239 | 98.5 | 569.8 | 5.4 | 127 | 569.8 | 0 | 121.6% | 38.0% | 25% |
| 2 | U[10,15] | 30432 | 104.5 | 880 | 3.4 | 128 | 885 | -5 | 94.8% | 57.9% | 52% |
| 3 | U[15,20] | 33066 | 107 | 890 | 3.2 | 127 | 898 | -8 | 82.9% | 57.4% | 62% |
| 4 | U[20,30] | 36175 | 105.6 | 870 | 3.0 | 128 | 1207.1 | -337 | 83.7% | 60.0% | 97% |

### 4.4.3 Base Price

We vary the base price $r$ in [$95, $150] by $5, and generate 12 instances. Based on the result in Figure 4, the impact of base price on the system is straightforward. First, a higher base price will bring more profit to the company, as well as increase the selling price $R$. Also, $R$ is less than $r$ for all cases; because of that, the overall demand will increase as the actual price is lower than the suggested price, which attracts more orders. In practice, this marketing strategy is frequently used, as customers are willing to spend money on a discounted product, and ignore the suggested price that is significantly higher.

### 4.4.4 Unit Production Time

Figure 5 shows how unit production time a makes an impact on profit, $Q$, and $D$. Clearly, both profit and $Q$ decrease as we raise a from 0.1 to 0.5. After all, a higher a indicates a lower productivity and capacity of the company, meaning that more demands can not be fulfilled, and that more shortage will occur, as indicated by a larger gap between $D$ and $Q$.

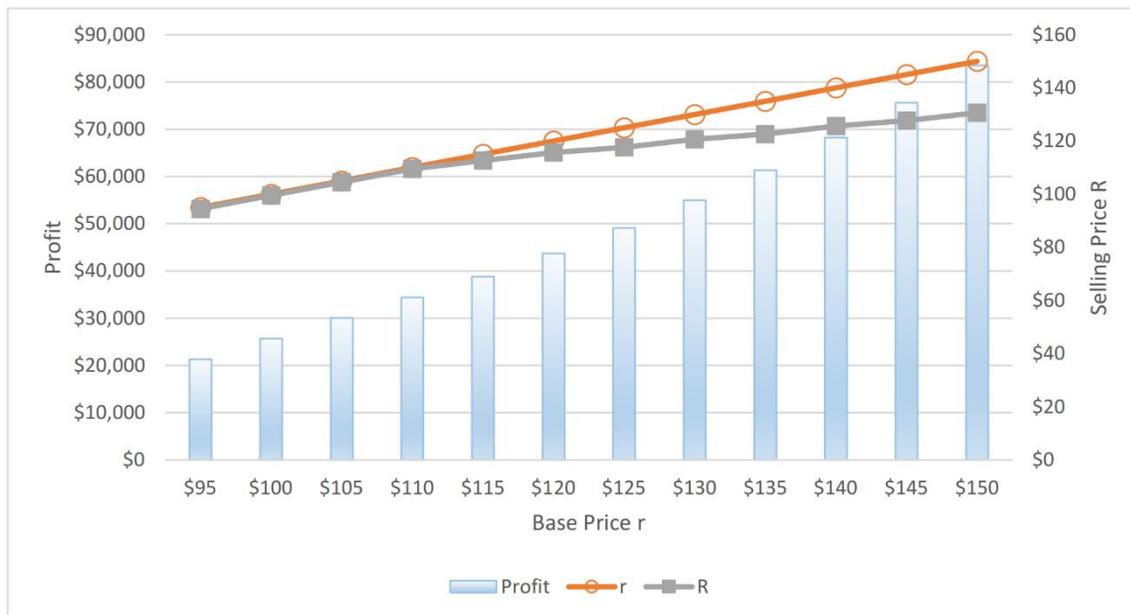

Figure 4: Impact of Base Price on Instance ($I = 4$, $J = 50$)

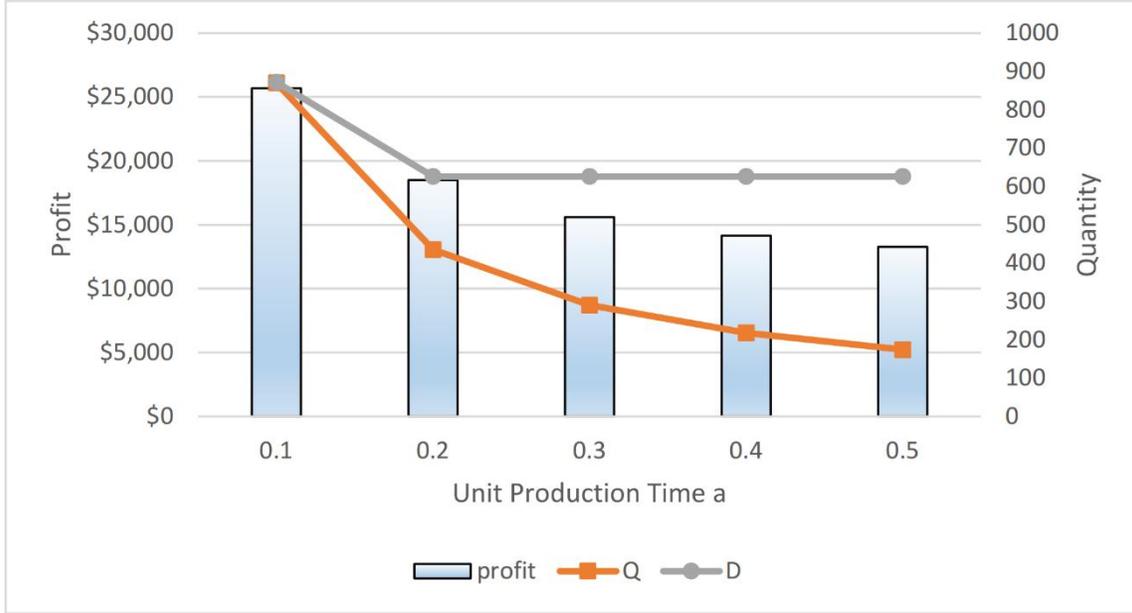

Figure 5: Impact of Unit Production Time on Instance ($I = 4$, $J = 50$)

Table 7: Impact of Price Scale $\lambda$ on Instance ($I = 4$, $J = 50$)

| ID | $\lambda$ | Profit | $R^*$ | $Q^*$ | T | $R_{ub}$ | D | $\Delta_{Q^*,D}$ | $M_1$ | $M_2$ | $M_3$ |
|----|-----|---------|-------|-------|------|------|-------|-------|--------|--------|------|
| 1  | 0.1 | 50180   | 170   | 501.8 | 19.0 | 229  | 501.8 | 0     | 66.4%  | 68.6%  | 100% |
| 2  | 0.2 | 33166   | 127.5 | 576.8 | 9.7  | 164  | 576.8 | 0     | 77.2%  | 78.8%  | 100% |
| 3  | 0.3 | 28244.6 | 113.5 | 649.3 | 6.0  | 143  | 649.3 | 0     | 87.6%  | 88.7%  | 100% |
| 4  | 0.4 | 26345.7 | 106.5 | 721.8 | 4.6  | 132  | 721.8 | 0     | 98.0%  | 98.6%  | 100% |
| 5  | 0.5 | 25657.6 | 102   | 801.8 | 3.6  | 126  | 801.8 | 0     | 109.4% | 109.5% | 100% |
| 6  | 0.6 | 25570.6 | 99.5  | 866.8 | 3.2  | 121  | 866.8 | 0     | 118.8% | 118.4% | 100% |
| 7  | 0.7 | 25644.4 | 99.5  | 869.3 | 2.5  | 118  | 869.3 | 0     | 119.1% | 118.8% | 100% |
| 8  | 0.8 | 25682.1 | 99.5  | 870   | 2.5  | 116  | 871.8 | -1.8  | 119.5% | 118.9% | 100% |
| 9  | 0.9 | 25705.8 | 99.5  | 870   | 2.2  | 114  | 874.3 | -4.3  | 119.8% | 118.9% | 100% |
| 10 | 1   | 25729.6 | 99.5  | 870   | 2.1  | 113  | 876.8 | -6.8  | 120.2% | 118.9% | 100% |
| 11 | 1.1 | 25753.3 | 99.5  | 870   | 1.7  | 111  | 879.3 | -9.3  | 120.5% | 118.9% | 100% |
| 12 | 1.2 | 25777.1 | 99.5  | 870   | 1.6  | 110  | 881.8 | -11.8 | 120.9% | 118.9% | 100% |
| 13 | 1.3 | 25800.8 | 99.5  | 870   | 1.6  | 110  | 884.3 | -14.3 | 121.3% | 118.9% | 100% |
| 14 | 1.4 | 25824.6 | 99.5  | 870   | 1.4  | 109  | 886.8 | -16.8 | 121.6% | 118.9% | 100% |

Table 8: Impact of Waiting Time on Instance ($I = 4$, $J = 50$)

| $w_j$ Range | Profit | $R^*$ | $Q^*$ | T | $R_{ub}$ | D | $\Delta_{Q^*,D}$ | $M_1$ | $M_2$ | $M_3$ |
|---|---|---|---|---|---|---|---|---|---|---|
| (20,40) | 12871 | 105 | 170 | 13.0 | 128 | 631.4 | -461.4 | 84% | 23% | 100% |
| (20,60) | 19090 | 112 | 310 | 26.6 | 140 | 585.9 | -275.9 | 73% | 31% | 68% |
| (20,80) | 29904 | 120 | 510 | 26.7 | 154 | 656.8 | -146.8 | 68% | 39% | 56% |
| (20,100) | 38540 | 121 | 470 | 29.7 | 167 | 940 | -470 | 65% | 29% | 78% |

### 4.4.5 Price Scale

The price scale $\lambda$ determines how much impact the price $R$ has on a customer's demand. In particular, $\lambda = 1$ means that the price increase/decrease by \$1, the demand will decrease/increase by 1 accordingly. The lower $\lambda$ is, the less impact it has on demand. Based on the data setting, we vary $\lambda$ from 0.1 to 1.4 by 0.1, and present the results in Table 7. As shown, when $\lambda$ is low, demand is insensitive to the selling price. Thus, the company can maximize its profit with a higher selling price. On the other hand, when $\lambda$ becomes higher, demand also becomes more sensitive to the price, forcing the company to reduce the selling price to gain a larger profit. In addition, we did not observe significant shortage in this data set, but a larger shortage is anticipated when $\lambda$ is higher, because the overall demand grows in a faster way.

### 4.4.6 Waiting Time

Table 8 demonstrates the impact of waiting time. For this experiment, we assume that waiting time is linearly related to expected demand. Letting $w_j = 2\mu_j$, we generate four instances by choosing $\mu_j$ randomly from [10, 20], [10, 30], [10, 40], and [10, 50], respectively. Thus, the ranges of $w_j$ are in [20, 40], [20, 60], [20, 80], and [20, 100] for the four tested instances. Also, to eliminate the influence of the service level constraint, we let $\alpha = 0$ for this experiment. The result shows that as the range of waiting times becomes larger, the solver eventually decides not to serve certain customers with short waiting times or low expected demand, because a short waiting time places a strong constraint on the total order quantity; thus it is likely that the company would rather not serve this customer, and choose other customers with longer waiting times and higher expected demand to generate more profit. In addition, we find that our solver takes a longer time (an average of 24 seconds) to finish an instance, as CPLEX takes a longer time to solve each MILP problem that needs more computation on customer selection. Column $M_3$ displays the ratio of served customers, in which three out of four instances have an actual service level lower than 80%. If the pre-set service level is 0.8, the company has to give up some customers to meet the required service level by sacrificing some profit.

### 4.4.7 Unit Production Cost

The impact of unit cost $c$ on the problem is obvious; as it only affects the total production cost in the whole problem, a higher unit cost will only reduce the profit by lowering the total production cost, i.e., the $cQ$ term in the objective function. There is no other impact of $c$ on the optimal solution.

## 5 Conclusion

In this paper, we define two new models, AON-SNP and DL-SNP, for the selective newsvendor problem. The first model considers an all-or-nothing demand pattern, while the second one takes into account a dependent lead time and a demand that can be affected by marketing factors like the selling price and customers' sensitivity to the price. The proposed two models address two different demand patterns which can both occur in the real world.

We develop the R-search algorithm to find an optimal solution for the DL-SNP problem. The R-search algorithm is based off a sequential searching on the selling price $R$, combined with mathematically proven optimizing strategies for performance improvement.

We conduct a series of experiments to evaluate the model. The result shows a premium performance of the R-search algorithm, in comparison with a sequential searching. In addition, we examine various responses of the system through parameter sensitivity analysis. Some key findings include:

- The selling price $R$ plays a crucial role in the model. First, it directly affects the demand. Second, it determines whether a customer should be selected to for service or not. It is not a trivial decision to decide a selling price and related marketing strategy such as the best discount rate. Our model can identify the best price that benefits the company most.

- The order quantity $Q$ could also be a tricky decision to make. Our model facilitates this decision. First, the dedicated supplier of the company has an upper limit of order quantity due to dependent lead time. Our model proves that if the total demand is lower than the upper limit of the order quantity, the best strategy for $Q$ is to match that demand; otherwise, if the market increases, more demand comes in, leading to a shortage, and forcing the company to find other local suppliers to fill the additional demand. However, it might not be a good idea to fill as much demand as possible. Our model can find the best shortage quantity that maximizes the profit.

- Customer satisfaction reflects the quality of service of a company. For our model, it is jointly affected by two metrics: $M_1$ and $M_3$. $M_1$ depends on the expected demand. The higher the expected demand is, the larger the potential market is. If the expected demand is accurate, the company has two strategies to respond: the company can either increase its productivity by lowering the unit production time, or outsource partial orders to another supplier. On the other hand, $M_3$ measures the fraction of selected customers over all customers. Clearly, $M_3$ should be as high as possible since denying a customer might hurt the reputation of a company.